\documentclass{ws-procs9x6}

\begin{document}
\title{Best Fit Predictions \\ for Flavour Violating Processes \\
       in MSSM Constrained by Unification\footnote{
                      \uppercase{T}alk presented by 
                      \uppercase{T}om\' a\v s
                      \uppercase{B}la\v zek.
  at {\it \uppercase{SUSY} 2003:
  \uppercase{S}upersymmetry in the \uppercase{D}esert}\/, 
  held at the \uppercase{U}niversity of \uppercase{A}rizona,
  \uppercase{T}ucson, \uppercase{AZ}, \uppercase{J}une 5-10, 2003.
  \uppercase{T}o appear in the \uppercase{P}roceedings.}}

\author{T.~BLA\v ZEK, S.~F.~KING AND J.~K.~PARRY}

\address{School of Physics and Astronomy \\
         University of Southampton\\ 
         Southampton, SO17 1BJ\\ 
E-mail: blazek@soton.ac.uk, sfk@soton.ac.uk, d63cf4@hep.phys.soton.ac.uk}


\maketitle

\abstracts{
We study the Minimal Supersymmetric Standard Model (MSSM) 
constrained by unification assuming mSUGRA SUSY breaking. In particular 
this means 
that the three gauge couplings are unified at a {\it per cent} level,
third generation yukawa couplings are unified at a $10\,$\% level,
and at the unification scale $M_U\approx 10^{16}$
the SUSY breaking masses of squarks
and sleptons are universal up to specific D-term contributions.
At $M_U$ we also include right-handed neutrinos that are  
decoupled at their three respective scales.
%
%
In a top-down global analysis we then find that there are
two distinct best fits, primarily distinguished by the
masses of the non-Standard Model Higgs states, and compute
the predictions for flavour violating processes involving
both leptons and quarks. The fit with the light Higgs spectrum
is especially interesting with regards to the decay
$B_s\to \mu^+\mu^-$. We note that
the latter results go beyond what is called minimal flavour
violation in the MSSM.
}

Within the diversity of different approaches to the flavour problem 
a class of supersymmetric (SUSY) unification models can be 
recognised that is remarkably simple at the unification scale. 
This is the class of SO(10)-like models where 
the Standard Model (SM) gauge couplings unify to a per cent 
level, third family yukawa couplings are all of
order unity, and the remaining flavour structure originates 
in a small set of higher-dimensional superpotential 
operators keeping the SUSY breaking sector flavour blind. 

\begin{figure}[ht]
\centerline{\epsfxsize=2.5in\epsfbox{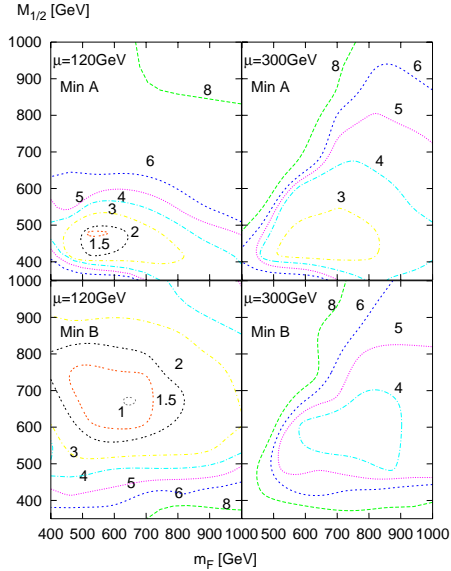}}   
\caption{$\chi^2$ contour plots in the plane of $(m_F,M_{1/2})$.  
         The four plots are obtained from the two minima, 
         Minimum A and Minimum B, with $\mu=120$ and $300$~GeV as labeled.  
         \label{pic_chi2}}
\end{figure}

Here we report on flavour violation \cite{bkp1,bkp2} in a Pati-Salam model
which falls into this category. Our work is general and applies 
to any mSUGRA model with universal sfermion masses at $M_U$ and a
Higgs spectrum similar to the one considered here.
In such a framework it is well known that, for a given choice
of low energy fermion masses and CKM mixing angles,
predictions on flavour violation in the quark sector 
are independent of the precise nature of the Yukawa
matrices selected at high energy since different choices of Yukawa
matrices can be rotated into each other. Apart from being general,
our study contains a number of new features not present
in previous works on flavour violating processes.
One novel feature is to present results that are 
based on a top-down global analysis. 
In a top-down approach 
complete Yukawa and sfermion mass matrices are known at the 
low energy scale and no extra iteration (called resummation
of large tan$\beta$ terms by some) is needed to extract the couplings which
enter the evaluation of the SUSY loops. Another new aspect 
is a more complete analysis of 
the process $B_s\to\mu^+\mu^-$. The previous analyses of SO(10)-like
models (see the list of references in Ref.~\refcite{bkp2}) 
only considered minimal flavour violation with the rate
$B_s\to\mu^+\mu^-$ explicitly proportional to the low-energy
value $V_{ts}^2$ while our results are more general and include 
contributions from additional diagrams.
Such contributions, which arise from inter-generational squark mixing 
effects, have so far been ignored in mSUGRA based analyses.

\begin{figure}[t]
\centerline{\epsfxsize=2.5in\epsfbox{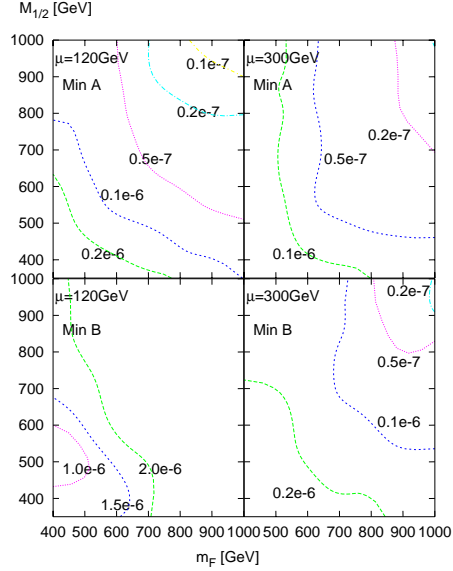}}   
\caption{Contours of $BR(\tau \rightarrow \mu \gamma)$ 
         plotted in the plane of $(m_F,M_{1/2})$.  
         The four plots, are obtained from the two minima, 
         Minimum A and Minimum B with $\mu=120$ and $300$~GeV 
         as labeled.  
         \label{pic_tmg}}
\end{figure}

In our top-down numerical analysis we vary the following gauge and SUSY 
parameters at the unification scale:      
               $M_{U},\;\alpha_{U},\;
                \epsilon_3\equiv(\alpha_3-\alpha_U)/\alpha_U$, 
     $m_F^2\equiv m_0^2,\;M_{1/2},
                                  \; m_h^2\;\mbox{and}\;D^2$.
$A_0(M_U)=0$ and $\tan\beta(M_{top})=50$  are kept fixed while 
$\mu(M_{top})$ can be controlled.
We also vary parameters of order unity that enter the hierarchical yukawa
and right-handed neutrino matrices at scale $M_U$. At low scale 
the one-loop effective potential is minimised and with the obtained
Higgs {\it vev}'s the  
$\chi^2$ is computed based on the fit of 24 observables. The latter include
the gauge couplings, fermion masses and mixings and vector boson masses.
We emphasise that the 32 flavour mixing is restricted 
by observables $V_{cb}$, \cite{bpr} $\:U_{\mu 3}$, and the $BR(b\to s \gamma)$.
Finally, constraints on unobserved sparticle masses are imposed.

\begin{figure}[t]
\centerline{\epsfxsize=2.5in\epsfbox{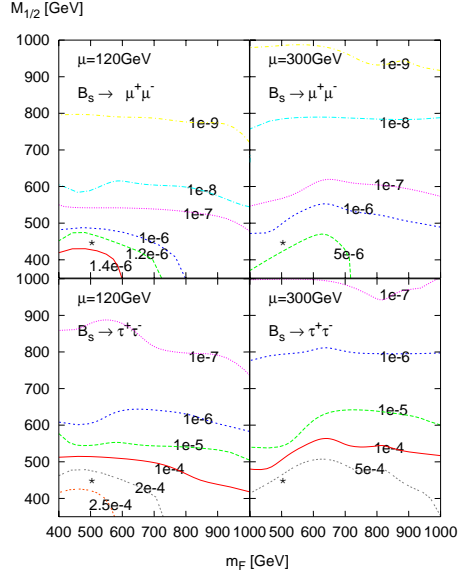}}   
\caption{Contour plots for the branching ratios of the FCNC processes,
         $B_{s}\to\mu^{+}\mu^{-}$ and $B_{s}\to\tau^{+}\tau^{-}$ 
         in Minimum A. Each branching ratio is plotted with 
         two different values of the $\mu$ parameter. 
         The $\star$ marks the Minimum A best fit point.
         \label{pic_Bs}}
\end{figure}

Two distinct minima, Minimum A and Minimum B, 
are found (see Figure \ref{pic_chi2}). They both predict 
$m_h\approx 115\,$GeV \cite{bdr}
and a low value 
for $\mu(M_{top})$ giving light higgsino-like chargino and neutralino.
The difference is the light (heavy) Higgs sector in Minimum A 
(Minimum B) and heavier (lighter) slepton masses in Minimum A
(Minimum B) --- see Figures 10 and 3 in the second reference \refcite{bkp1}. 
The latter result leads to a significant prediction for 
the lepton flavour violating $\tau\to \mu\gamma$ decay, 
Figure \ref{pic_tmg}, at the present limit (Minimum B) 
or within the reach of the B factories in the near future (Minimum A).
More importantly, the light pseudoscalar Higgs in Minimum A leads to
the exciting prediction for Higgs mediated pure leptonic 
$B_s\to\mu^+\mu^-$ decay (the two upper panels in Figure \ref{pic_Bs}).
Sensitivity to $BR(B_s\to\mu^+\mu^-)\approx 10^{-7}$ at the Tevatron 
would be very restrictive and could probe 
$m_{A^0}\stackrel{<}{\approx}300\,$GeV, a truly remarkable result.
We repeat that it has been derived based on the full computation in 
generation space that goes beyond the minimal flavour violation.
In Minimum B, $B_s\to\mu^+\mu^-$ is not enhanced over the small SM rate
due to the much heavier Higgs spectrum.


\end{document}